\documentclass[amsmath, amsfonts, showpacs, superscriptaddress, twocolumn, prb]{revtex4}
\usepackage{graphicx}
\usepackage{epsfig}
\usepackage{bm}
\usepackage{dcolumn}
\usepackage{amsmath}
\usepackage{amssymb}

\begin{document}

\title{Thermal transport and quench relaxation in nonlinear Luttinger liquids}

\author{Stanislav Apostolov}

\affiliation{Department of Physics and Astronomy, Michigan State
University, East Lansing, Michigan 48824, USA}

\affiliation{A.Ya. Usikov Institute for Radiophysics and
Electronics, National Academy of Sciences of Ukraine, 61085 Kharkov,
Ukraine}

\author{Dong E.~Liu}
\affiliation{Department of Physics and Astronomy, Michigan State
University, East Lansing, Michigan 48824, USA}

\author{Zakhar Maizelis}

\affiliation{Department of Physics and Astronomy, Michigan State
University, East Lansing, Michigan 48824, USA}

\affiliation{A.Ya. Usikov Institute for Radiophysics and
Electronics, National Academy of Sciences of Ukraine, 61085 Kharkov,
Ukraine}

\author{Alex Levchenko}
\affiliation{Department of Physics and Astronomy, Michigan State
University, East Lansing, Michigan 48824, USA}

\begin{abstract}
One-dimensional electrons with a linearized dispersion relation are
equivalent to a collection of harmonic plasmon modes, which
represent long wavelength density oscillations. An immediate
consequence of this Luttinger model of one-dimensional electron
systems is the absence of inelastic scattering processes responsible
for the relaxation of nonequilibrium states. In a generic nonlinear
Luttinger liquid plasmons may decay and thus acquire a finite
lifetime. We show that equilibration of plasmons is hierarchical and
has profound implications for the dynamics after a thermal quench.
We also develop a thermal transport theory and compute thermal
conductance of the nonlinear Luttinger liquid by treating the
collision integral of plasmons in a manifestly nonperturbative way.
\end{abstract}

\date{July 11, 2013}

\pacs{71.10.Pm, 72.10.-d, 72.15.Lh, 73.20.Qt, 73.21.Hb}

\maketitle

\section{Introduction}

The kinetics of nearly integrable quantum many-body systems is the
subject of ongoing study. One-dimensional (1D) systems are special
in this regard since several exact solutions are
known,~\cite{Mattis,Sutherland} and perhaps using them for generic
models where integrability is broken weakly makes a good starting
point. Integrability ensures that scattering of the particles of an
$N$-body system is exactly equivalent to a sequence of pair-particle
collisions, and thus the set of incoming momenta for any scattering
event coincides with the set of outgoing momenta. Such
non-diffractive scattering does not alter the distribution function
and is unable to drive the system towards thermal equilibrium. A
quantum Newton's cradle, realized with trapped 1D Bose gas, gives
the best example for such long-lived out-of-equilibrium quantum
states unaffected by binary collisions.~\cite{Kinoshita}

The exactly solvable Tomonaga-Luttinger (TL)
model~\cite{Tomonaga,Luttinger} provided a framework for the study
of 1D electron liquids as realized in quantum wires, nanotubes and
edge states, see Ref.~\onlinecite{Deshpande} for a recent review and
comprehensive list of references. This model was extremely
successful in predicting peculiar properties of such Luttinger
liquids (LL), most notable ones being power-law anomalies in the
tunneling density of states,~\cite{Kim-ZBA} and effect of
spin-charge separation.~\cite{Auslaender} However, it also possesses
some serious deficiencies. Within the Luttinger model excitations
have infinite lifetime, which implies the lack of equilibration.
Interactions in the perfect Luttinger liquid conductor do not affect
two-terminal conductance as compared to its noninteracting value
$\mathcal{G}_0=2e^2/h$. Due to built-in particle-hole symmetry TL
model misses completely effects such as thermopower, photo-voltaic
response etc.

The resurgence of interests in 1D electron liquids is triggered by
experimental results which clearly fall outside of the LL paradigm.
Energy and spatially-resolved tunneling spectroscopy with quantum
wires~\cite{Chen,Barak} and local thermometry with driven quantum
Hall edge states~\cite{Granger,Altimiras} provided direct evidence
for the electronic thermalization in 1D systems. Transport
measurements in low density wires revealed deviations from perfect
conductance quantization~\cite{G1,G2,G3,G4,G5} and violation of the
Wiedemann-Franz (WF) law.~\cite{Chiatti,Wakeham} These observations
attracted much theoretical attention and brought to the agenda the
new concept of nonlinear Luttinger liquids.~\cite{Adilet} In the
context of the recent studies the subjects of our interest are (i)
the microscopic mechanisms of relaxation in the generalized TL
description of 1D electron liquids, which amounts to keeping
anharmonic interactions between plasmons, and (ii) the way such
liquids transfer energy.

Study of the thermal transport in LL was pioneered by Kane and
Fisher.~\cite{KF} They concluded that: (i) in pure LL the thermal
conductance $\mathcal{K}$ does not depend on interaction and
coincides with its noninteracting value $\mathcal{K}_0=2\pi^2T/3h$,
so that the Lorentz number $\mathcal{L}=\mathcal{K}/T\mathcal{G}$ is
still $\mathcal{L}_0=\pi^2/3e^2$ and the WF law holds. (ii) In the
presence of a single impurity strong electron backscattering
modifies both $\mathcal{G}$ and $\mathcal{K}$ so that
$\mathcal{L}/\mathcal{L}_0=3/(2\kappa+\kappa^2)$ for the Luttinger
interaction parameter $1/2<\kappa<1$, while the Lorentz number
diverges $\mathcal{L}\propto T^{4-2/\kappa}$ as $T\to0$ for
$\kappa<1/2$. Large violation of the WF law was also predicted for
disordered LL on the lattice when umklapp scattering rate exceeds
the impurity scattering.~\cite{Rosch} This problem was also analyzed
in Refs.~\onlinecite{FHK,Krive,GGM} within the model of
inhomogeneous LL with spatially dependent interaction parameter
$\kappa(x)$.~\cite{MS} If $\kappa(x)$ varies smoothly on the scale
of Fermi wave length then electrons do not suffer backscattering so
that conductance is still $2e^2/h$. However, plasmons representing
low energy excitation of the LL do backscatter which strongly
renormalizes $\mathcal{K}$. Physically this effect can be traced
back to the inhomogeneity-induced lifetime of plasmons.~\cite{GR}
Note that anharmonic terms neglected within TL model also lead to
interaction between plasmons. The effect of these two scattering
mechanisms on the plasmon kinetics is very different. Indeed, in the
system that lacks translational invariance momentum can be relaxed
by scattering off inhomogeneities but taken alone does not lead to
thermal equilibrium. In what follows we calculate thermal
conductance of the genuine (homogeneous) nonlinear LL accounting for
the inelastic scattering of plasmons.

\section{Model}

The simplest way to visualize charge and heat transport in
repulsively interacting electron system is to consider the motion of
1D Wigner crystal (WC)~\cite{Schulz,Matveev-WC} through the
constriction. Wigner crystal, as realized in carbon
nanotubes~\cite{Bockrath-NatPhys08} and quantum
wires~\cite{Ritchie-PRL09,Yamamoto-PRB12}, represents an extreme
case of the LL with interaction parameter
$\kappa=\pi\hbar\rho^2/ms\ll1$, where $\rho$ is particle density,
$m$ is electron mass and $s$ is sound velocity of plasmons. At zero
temperature the rigid shift of WC results in interaction independent
conductance $\mathcal{G}_0$. It is evident that at finite
temperatures thermally activated plasmon waves in WC can not affect
electrical transport through perfect LL constriction. In fact such
corrections require umklapp scattering which are exponentially
suppressed at low temperatures and thus can be safely disregarded.
However scattering of plasmons may easily alter energy transfer.

We follow recent work~\cite{Matveev-WC-Eq} and model the system of
strongly interacting spinless electrons by the Hamiltonian
(hereafter $\hbar=1$)
\begin{equation}\label{H}
H=\sum_l\frac{p^2_l}{2m}+\frac{1}{2}\sum_{l\neq l'}V(x_l-x_{l'})
\end{equation}
where $p_l$ and $x_l$ are the momentum and coordinate of the $l$th
particle ($l=1,\ldots,N$), and $V(x)$ is the interaction potential.
In calculations we assume screened Coulomb potential
$V(x)=\frac{e^2}{\epsilon}\big[\frac{1}{|x|}-\frac{1}{\sqrt{x^2+4d^2}}\big]$,
where $d$ is the distance to the screening gate and $\epsilon$ is
the dielectric constant of the host material of the wire. For this
model the WC state exists only in the density range
$a_B/d^2\ll\rho\ll a^{-1}_B$,~\cite{Matveev-WC} where
$a_B=\epsilon/me^2$ is the effective Bohr radius of the material. In
the WC picture electrons form a lattice while deviations
$u_l=x_l-l/\rho$ from the corresponding equilibrium sites describe
the low energy excitations in the form of electron density waves.
Indeed, assuming that relative change of interparticle distance
remains small, $|u_l-u_{l'}|\ll|l-l'|/\rho$, expanding Eq.~\eqref{H}
to the second order in $u_l$, and passing to the oscillator
representation with
$p_l=-i\sum_q\sqrt{\frac{m\omega_q}{2N}}(b_q-b^\dag_{-q})e^{iql}$
and
$u_l=\sum_q\sqrt{\frac{1}{2mN\omega_q}}(b_q+b^\dag_{-q})e^{iql}$,
one finds that quadratic part of Eq.~\eqref{H} takes the usual LL
form $H_0=\sum_q\omega_q(b^\dag_qb_q+1/2)$, where operators
$b^\dag_q(b_q)$ conventionally created (annihilate) one boson. The
plasmon dispersion is given by
$\omega^2_q=(2/m)\sum_lV^{(2)}_l[1-\cos(ql)]$, where
$V^{(n)}_{l}=\partial^{n}_xV(x)|_{x=l/\rho}$, which reduces to
$\omega_q=s|q|$ in the low energy limit $q\to0$. We choose to
measure all momenta in units of density so that plasmon velocity
$s=(e^2\rho/\varepsilon)\sqrt{2\rho a_B\ln(\rho d)}$ has units of
energy.

Anharmonic terms in the expansion of Eq.~\eqref{H} in powers of
$u_l$ lead to interaction of plasmons and thus manifestly break
integrability of the model. Quartic nonlinearity and qubic terms
iterated to the second order generate two-plasmon scattering
process. Ignoring the possibility of coherence between different
particle states, i.e. off diagonal elements of the density matrix,
plasmons can be described by the distribution function
$\mathcal{N}(q,x,t)$, which obeys Boltzmann kinetic equation (BKE)
with the collision integral
\begin{eqnarray}\label{Coll-Int-1}
I[\mathcal{N}_1]=-\sum_{q_2q'_1q'_2}W_{QQ'}[\mathcal{N}_1\mathcal{N}_2
(1+\mathcal{N}_{1'})(1+\mathcal{N}_{2'})\nonumber\\
-\mathcal{N}_{1'}\mathcal{N}_{2'}(1+\mathcal{N}_1)(1+\mathcal{N}_2)]
\end{eqnarray}
where we used shorthand notation
$\mathcal{N}_i=\mathcal{N}(q_i,x,t)$. The scattering rate
$W_{QQ'}=2\pi|A_{QQ'}|^2\delta_{Q,Q'}\delta(\Omega-\Omega')$ follows
from the Fermi's golden rule with the amplitude:
$|A_{QQ'}|^2=\big(\frac{\lambda\rho^2}{mN}\big)^2|q_1q_2q'_1q'_2|$.~\cite{Matveev-WC-Eq}
Conventionally, the delta-functions in $W_{QQ'}$ account for the
momentum and energy conservations, with $Q=q_1+q_2$,
$\Omega=\omega_{q_1}+\omega_{q_2}$, and similarly for the primed
values in the final state. The dimensionless number $\lambda$ is not
universal and depends on the choice of interaction potential $V(x)$.
For the special potentials that correspond to exactly solvable
models~\cite{Sutherland} it vanishes identically, while
$\lambda=-3/4$ for the screened Coulomb case.~\cite{Matveev-WC-Eq}
In order to arrive to that particular form of $A_{QQ'}$ one
necessarily has to account for the nonlinearity of the plasmon
dispersion $\omega_q$ in order to regularize otherwise diverging
scattering rate. Collisions encoded by Eq.~\eqref{Coll-Int-1}
kinematically allow a process when two co-moving plasmons scatters
into two counter-propagating ones. Such scattering results in the
momentum and energy transfer between the left and right branches of
the spectrum, and leads to thermalization.

\section{Thermal conductance}

Consider now WC quantum wire of length $\ell$ subject to a thermal
bias. We assume that right(left) lead is kept at temperature
$T_{r(l)}$ such that $T_r-T_l=\Delta T$. In order to find thermal
conductance in such setting we need to calculate energy current
carried by plasmons. The latter requires the knowledge of the
nonequilibrium plasmon distribution function along the wire, which
should follow from the solution of the BKE in a steady state
\begin{equation}\label{BKE}
s_q\partial_x \mathcal{N}(q,x)=I[\mathcal{N}(q,x)],
\end{equation}
were $s_q=\partial_q\omega_q$. Needless to say that finding solution
of the integro-differential equation \eqref{BKE} even within the
linear response to $\Delta T$ is a complicated problem. We achieve
this goal by adopting an approach recently developed for the same
problem in the case of weakly interaction 1D electron
gas.~\cite{AL-K,TM} To this end, it is convenient to parameterize
distribution function of plasmons in the wire as follows
\begin{equation}\label{N}
\mathcal{N}(q,x)=\frac{\theta(q)}{e^{(\omega_q-u^Rq)/T^R}-1}
+\frac{\theta(-q)}{e^{(\omega_q-u^Lq)/T^L}-1}.
\end{equation}
Here $\theta(\pm q)$ is the step-function, upper indices $R(L)$
refer to right(left) moving plasmons. Parameters $u^{R(L)}$ have
meaning of plasmon drift velocities while $T^{R(L)}$ are their
effective temperatures. The ansatz made by Eq.~\eqref{N} can be
understood as follows. The scattering process responsible for the
energy exchange between plasmons involves three plasmon momenta
$q_1,q_2,q'_1$ that belong to the same branch (say, right-moving),
while forth one $q_{2'}$ has to have opposite sign as dictated by
the energy and momentum conservation laws. Furthermore, at low
temperatures the characteristic scale of $q'_2\sim (T/s)^3$ is
parametrically smaller than that of $\{q_1,q_2,q'_1\}\sim T/s$. This
implies that redistribution of energy between co-moving plasmons
occurs much more efficiently than energy transfer to
counter-propagating branch. With this observation at hand, the
partially equilibrated form of the plasmon distribution function
\eqref{N} can be obtained from a general statistical mechanics
argument by maximizing the entropy
$S=\sum_q[(\mathcal{N}+1)\ln(\mathcal{N}+1)-\mathcal{N}\ln\mathcal{N}]$,
under the constraints of approximately conserved quantities. The
Lagrange multipliers $u^{R(L)}(x)$ and $T^{R(L)}(x)$ are functions
of coordinate along the wire that we need to determine. For this
purpose we introduce momentum $j_P$ and energy $j_E$ currents of
plasmons
\begin{equation}\label{j-P-E}
\left\{\begin{array}{c} j^{R(L)}_{P}(x) \\ j^{R(L)}_{E}(x)
\end{array}\right\}=\int^{+\infty}_{-\infty}\frac{dq}{2\pi}\theta(\pm
q)s_q\left\{\begin{array}{c}q \\ \omega_q
\end{array}\right\}
[\mathcal{N}(q,x)-n_q]
\end{equation}
where $n_q=(e^{\omega_q/T}-1)^{-1}$ is equilibrium Bose distribution
function. First, we observe that total currents are protected by
conservation laws and thus must obey continuity equations
\begin{equation}\label{Eq1}
\partial_x [j^R_P(x)+j^L_P(x)]=0,\qquad \partial_x[j^R_E(x)+j^L_E(x)]=0.
\end{equation}
Second, the corresponding currents of right(left)-moving plasmons
are not independently conserved and follow from the kinetic
equations
\begin{equation}\label{Eq2}
\partial_x[j^R_P(x)-j^L_P(x)]=2\dot{p}^R,\quad
\partial_x[j^R_E(x)-j^L_E(x)]=2\dot{\varepsilon}^R
\end{equation}
where quantities $\dot{p}^R$ and $\dot{\varepsilon}^R$ are
respective rates of momentum and energy change, which have to be
computed from the collision integral Eq.~\eqref{Coll-Int-1}. Within
the linear response to applied thermal bias $\Delta T$ we expand
$\mathcal{N}(q,x)$ in Eq.~\eqref{N} to the leading order in
$u^{R(L)}(x)$ and $\delta T^{R(L)}(x)=T^{R(L)}(x)-T$, and calculate
currents from Eq.~\eqref{j-P-E}. When combined with Eqs.~\eqref{Eq1}
and \eqref{Eq2} this gives us a system of four coupled linear first
order differential equations that govern spatial evolution of the
Lagrange multipliers that parameterize $\mathcal{N}(q,x)$ in
Eq.~\eqref{N}. Specifically we find
\begin{eqnarray}
&&\hskip-.8cm g_1\partial_x\vartheta_++g_2\partial_x\eta_-=0,\quad
\partial_x\vartheta_-+g_1\partial_x\eta_+=0,\\
&&\hskip-.8cm
g_1\partial_x\vartheta_-+g_2\partial_x\eta_+=-\frac{\eta_-}{\ell_E},\quad
\partial_x\vartheta_++g_1\partial_x\eta_-=-\frac{\vartheta_-}{\ell_E}.
\end{eqnarray}
Here first two equations correspond to the conservation laws in
Eq.~\eqref{Eq1}, while the last two correspond to the kinetic
equations in Eq.~\eqref{Eq2}. We introduced notations
$\vartheta_\pm=(\delta T^R\pm\delta T^L)/T$ and $\eta_\pm=(u^R\pm
u^L)/s$, and two functions
$g_1=1+\frac{\xi\tau^2}{5}+\frac{\xi^2\tau^4}{2}$ and
$g_2=1+\frac{2\xi\tau^2}{5}+\frac{8\xi^2\tau^4}{7}$, where
$\tau=2\pi T/s$ and $\xi=\frac{(\rho d)^2\ln\tau}{2\ln(\rho d)}$.
Keeping the nonlinear terms in the plasmon dispersion $\omega_q$ and
carrying low-temperature expansion $\tau\ll1$ is justified in the
limit of strong interaction provided that $\kappa\ll(\xi\tau)^2$.
The relaxation length $\ell_E$ in Eqs.(8-9) is given by
\begin{equation}
\ell^{-1}_{E}=\frac{6}{\pi NT^3} \sum_{q_1>0q_2>0\atop
q'_1>0q'_2<0}\omega^2_{q'_{2}}W_{QQ'}n_{q_1}n_{q_2}(1+n_{q'_{1}})(1+n_{q'_{2}})
\end{equation}
To solve Eqs.~(8-9) we need to supplement them by the proper
boundary conditions. To find the latter we notice that once electron
enters the interacting part of the wire from the leads it breaks up
into plasmons. Given its excess energy, determined by the thermal
bias $\Delta T$ in our case, conservation laws uniquely determine
energy partitioning between right- and left-moving created
plasmons.~\cite{GGM,Karzig} We thus find
\begin{eqnarray}
&&u^R(0)=-\mathcal{R}u^L(0),\quad u^L(\ell)=-\mathcal{R}u^R(\ell),\\
&&\delta T^R(0)=\mathcal{T}\Delta T/2+\mathcal{R}\delta T^L(0),\\
&&\delta T^L(\ell)=-\mathcal{T}\Delta T/2+\mathcal{R}\delta
T^R(\ell).
\end{eqnarray}
The transmission $\mathcal{T}$ and reflection $\mathcal{R}$
coefficients of plasmons from the boundaries separating interacting
and noninteracting parts of a wire are given by the Fresnel law:
$\mathcal{T}=4\kappa/(1+\kappa)^2$ and
$\mathcal{R}=(1-\kappa)^2/(1+\kappa)^2$. Notice that $\delta
T^{R(L)}$ is not simply $\pm\Delta T/2$, which is essentially due to
the quantum analog of the Kapitza boundary thermal
resistance.~\cite{Kapitza-exp} This leads us to the final result for
the thermal conductance $\mathcal{K}=j_E/\Delta T$:
\begin{equation}\label{K}
\frac{\mathcal{K}}{\mathcal{K}_0}=\chi g_1 \frac{(1+\chi
g_1)+(1-\chi g_1)e^{-\ell/\ell_{pl}}} {(g_1+\chi g_2)+(g_1-\chi
g_2)e^{-\ell/\ell_{pl}}},
\end{equation}
where $\chi=\mathcal{T}/(1+\mathcal{R})$, and interaction-induced
inelastic scattering length of plasmons is given by~\cite{Note}
\begin{equation}\label{L-pl}
\ell^{-1}_{pl}=\frac{175s^4}{288\pi^4\xi^2T^4}\ell^{-1}_E=c_{\ell}\lambda^2\kappa^2(T/s)^5.
\end{equation}
It is instructive to analyze Eq.~\eqref{K} in various limiting
cases. For short wires, $\ell\ll\ell_{pl}$, the effect of plasmon
scattering is weak and interaction-induced correction
$\delta\mathcal{K}$ to thermal conductance can be obtained by
perturbation theory. It scales linearly with the length of the wire,
$\delta \mathcal{K}/\mathcal{K}_0\propto-\ell/\ell_{pl}$. In
contrast, for long wires $\ell\gg\ell_{pl}$, thermalization of
plasmons leads to saturation of conductance
$\mathcal{K}/\mathcal{K}_0=2\kappa\big[1-\frac{36\kappa\xi^2\tau^4}{175}\big]$,
where we approximated $\chi\approx2\kappa$ for the WC limit
$\kappa\ll1$. Conversely, neglecting thermalization effects but
assuming arbitrary interactions we recover from Eq.~\eqref{K}
$\mathcal{K}/\mathcal{K}_0=2\kappa/(1+\kappa^2)$.

We wish to notice that scattering length in Eq.~\eqref{L-pl} is
directly related to the plasmon lifetime $\tau_{pl}=\ell_{pl}/s$,
which is in agreement with the results of the recent study
Ref.~\onlinecite{Matveev-WC-Eq}. In addition, there is one relevant
detail which is worth emphasizing. Scattering process considered so
far conserves number of plasmons in the initial and final states.
This would imply that plasmons acquire chemical potential, which is
of course relaxed by the other inelastic processes. The most
relevant one involves two plasmons scattering into three or vise
versa, which is governed by the collision in integral
\begin{eqnarray}\label{Coll-Int-2}
I[\mathcal{N}_1]=&&-\!\!\!\!\!\!\!
\sum_{q_2q'_1q'_2q'_3}\!\!\!\widetilde{W}_{QQ'}
[\mathcal{N}_1\mathcal{N}_2
(1+\mathcal{N}_{1'})(1+\mathcal{N}_{2'})(1+\mathcal{N}_{3'})\nonumber\\
&&-\mathcal{N}_{1'}\mathcal{N}_{2'}\mathcal{N}_{3'}(1+\mathcal{N}_1)(1+\mathcal{N}_2)].
\end{eqnarray}
with the amplitude
$|\widetilde{A}_{QQ'}|^2=\frac{2\tilde{\lambda}^2\rho^6}{s(mN)^{3}}|q_1q_2q'_1q'_2q'_3|
$, where we find $\tilde{\lambda}=55/48$.~\cite{SM} Kinematics of
this process is such that all momenta transferred are of the same
order $\{q,q'\}\sim T/s$ and are on the same branch. Thus this
process is not relevant for the energy exchange and the thermal
transport, nevertheless it leads to a finite scattering length
$\tilde{\ell}^{-1}_{pl}\simeq(\tilde{\lambda}^2/\xi)\kappa^3(T/s)^5$.
Surprisingly, it has the same temperature dependence as
$\ell^{-1}_{pl}$ in Eq.~\eqref{L-pl}, while naively one would expect
to have it to the higher order in $T$ since amplitude and collision
integral contain extra powers of $q$. Two length scales are
parametrically distinct in interaction strength
$\tilde{\ell}^{-1}_{pl}/\ell^{-1}_{pl}\sim\kappa\ll1$.

\section{Thermal quench relaxation}

Up to this point we concentrated on the scattering of plasmon
excitations and their contribution to the thermal conductance in
generic LL. A related question, which attracted a lot of attention
recently, is the effect of interactions on the relaxation of sudden
quenches in LL.~\cite{Iucci,Mitra,Foster} What is still not
completely resolved is the problem of integrability-breaking
perturbations on the quench dynamics, decay of
currents~\cite{Achim,Sirker,Roni} and ultimately approach to thermal
equilibrium.~\cite{Polkovnikov} We address these issues in the
context of the quenched nonlinear LL by studying its evolution at
the time scales exceeding inelastic scattering time.

For the time dependent setting it has been shown in
Ref.~\onlinecite{Matveev-WC-Eq} that in the parametrization
$\mathcal{N}(q,t)=n_q+g_q\phi(q,t)$, where $g_q=\sqrt{n_q(1+n_q)}$,
linearized BKE for the plasmon collision integral
Eq.~\eqref{Coll-Int-1} can be reduced to the eigenvalue problem for
the following integro-differential equation
\begin{equation}\label{Coll-Int-3}
\partial_t\phi(p,t)=-\tau^{-1}_{pl}\int^{\infty}_{0}\mathbb{K}(p,p')\phi(p',t)dp'.
\end{equation}
The kernel is given by
$\mathbb{K}(p,p')=\frac{1}{6}p^2(p^2+1)\delta(p-p')+\frac{pp'(p+p')}{\sinh[\pi(p+p')]}-
\frac{pp'(p-p')}{\sinh[\pi(p-p')]}$ and dimensionless variable
$p=sq/2\pi T$ parametrizes momentum. Above integral operator has
continuous spectrum $\zeta_\nu=\nu^2(\nu^2+1)/6$ and can be
diagonalized in the basis of the eigenfunctions
$\psi_\nu(p)=\Delta^{-1}_\nu[(2\nu^2-1)\delta(p-\nu)+
\frac{3p}{\sinh[\pi(p+\nu)]}+\frac{3p}{\sinh[\pi(p-\nu)]}]$ where
the norm is $\Delta_\nu=\sqrt{(\nu^2+1)(4\nu^2+1)}$, such that
$\phi(p,t)=\int^{\infty}_{0}\alpha_\nu\psi_\nu(p)e^{-\zeta_\nu
t/\tau_{pl}}d\nu$. Expansion coefficients should be determined from
the initial conditions
$\alpha_\nu=\int^{\infty}_{0}\phi(p,0)\psi_\nu(p)dp$. Imagine now a
situation that excess energy is suddenly added to the electron
liquid which generates heat current. In the actual experiments this
could be realized either by selective tunneling of high energy
carriers into the quantum wire or by local Joule heating via low
conductance quantum point contacts. Relaxation than, in principle,
can be monitored by the local thermometry based on the
thermoelectric effect. We thus interested in calculating the decay
of the energy current carried by (say, right-moving) plasmons. Since
Eq.~\eqref{Coll-Int-3} was derived by linearizing plasmon dispersion
and furthermore neglecting momentum transferred to left-movers
$q'_2\to0$ the eigenmodes of Eq.~\eqref{Coll-Int-3} possess a
spurious property that $j^R_E$ is conserved. To overcome this issue
we calculate energy transfer rate $\dot{\varepsilon}^R$ directly
from the initial collision integral Eq.~\eqref{Coll-Int-1} in the
basis of the approximate eigenmodes $\psi_\nu(q)$ by keeping
transferred momentum explicitly
\begin{eqnarray}\label{E-R}
&&\hskip-.25cm
\dot{\varepsilon}^R(t)=\frac{1}{N}\!\!\!\!\sum_{q_1>0q_2>0\atop
q'_1>0q'_2<0}\!\!\!\!\omega_{q'_2}W_{QQ'}n_{q_1}n_{q_2}(1+n_{q'_1})(1+n_{q'_2})\\
&&\hskip-.25cm
[\phi(q_1,t)/g_{q_1}+\phi(q_2,t)/g_{q_2}-\phi(q'_1,t)/g_{q'_1}-\phi(q'_2,t)/g_{q'_2}].\nonumber
\end{eqnarray}
We choose initial condition assuming that right-moving plasmons are
hot $\phi^R(q,0)=g_q\omega_q\delta T^R(0)/T^2$. At long times,
$t\gg\tau_{pl}$, high-energy plasmons already equilibrated, however
thermal plasmons with characteristic momenta $q\sim T/s$ are not yet
and their eigenmode expansion at that limit can be approximated by
$\phi(p,t)\approx\phi(p,0)e^{-p^2t/6\tau_{pl}}-\frac{6p}{\sinh(\pi
p)}\int^{\infty}_{0}\phi(\nu,0)e^{-\nu^2t/6\tau_{pl}}d\nu$, which
amounts of taking $\nu\ll1$ limit of $\psi_\nu(p)$. Recalling that
$j^R_E\propto\dot{\varepsilon}^R$ one can readily extract long time
asymptote of the energy current flux from Eq.~\eqref{E-R}, which
when normalized to the total injected current reads
\begin{equation}
j^R_E(t)/j^R_E(0)\simeq(\tau_{pl}/t)^{3/2},
\end{equation}
and exhibits non-exponential decay. The power-exponent in $j^R_E(t)$
is not universal and depends on the initial condition, however, the
power-law dependence at long times is generic feature of the thermal
quench relaxation in nonlinear Luttinger liquids.

\section{Summary}

We have presented the study of the thermal transport in the generic
1D electron system beyond its integrability limit. Combined effects
of the energy partitioning and thermalization of plasmons modify
thermal conductance of the nonlinear LL in the nontrivial way. The
leading order relaxation process stems from the two-plasmon
collisions, however complete equilibration is achieved by inelastic
scattering that do not conserve number of plasmon modes.
Interestingly, both scattering rates have the same temperature
dependence $\propto T^5$. The same physics is responsible for the
thermal quench relaxation. We conclude that at the time scales
exceeding interaction-induced lifetime of plasmons thermal currents
display non-exponential decay, which can be rooted to the fact that
eigenvalues of the collision integral for the plasmon scattering
cover continuous spectrum of excitations.

\section*{Acknowledgment}

We would like to thank Norman Birge and Mark Dykman for reading and
commenting on the paper. This work was supported by Michigan State
University and in part by ARO through contract W911NF-12-1-0235.


\begin{thebibliography}{99}

\bibitem{Mattis}
D.~C.~Mattis, \textit{The Many-Body Problem: An Encyclopedia of
Exactly Solved Models in One Dimension}, World Scientific
Publishing, 1992).

\bibitem{Sutherland}
B.~Sutherland, \textit{Beautiful models: 70 years of exactly solved
quantum many-body problems}, (World Sci. Pub., 2004).

\bibitem{Kinoshita}
T.~Kinoshita, T.~Wenger, and D.~Weiss, Nature \textbf{440}, 900
(2006).

\bibitem{Tomonaga}
S.~Tomonaga, Prog. Theor. Phys. \textbf{5}, 544 (1950).

\bibitem{Luttinger}
J.~M.~Luttinger, J. Math. Phys. \textbf{4}, 1154 (1963).

\bibitem{Deshpande}
V.~V.~Deshpande, M.~Bockrath, L.~I.~Glazman, and A.~Yacoby, Nature
\textbf{464}, 209 (2010).

\bibitem{Kim-ZBA}
L.~Venkataraman, Y.~S.~Hong, and P.~Kim, Phys. Rev. Lett.
\textbf{96}, 076601 (2006).

\bibitem{Auslaender}
O.~M.~Auslaender, H.~Steinberg, A.~Yacoby, Y.~Tserkovnyak,
B.~I.~Halperin, K.~W.~Baldwin, L.~N.~Pfeiffer, and K.~W.~West,
Science \textbf{308}, 88 (2005).

\bibitem{Chen}
Y.-F.~Chen, T.~Dirks, G.~Al-Zoubi, N.~O.~Birge, and N.~Mason, Phys.
Rev. Lett. \textbf{102}, 036804 (2009).

\bibitem{Barak}
G.~Barak, H.~Steinberg, L.~N.~Pheiffer, K.~W.~West, L.~Glazman,
F.~von Oppen, and A.~Yacoby, Nat. Phys. \textbf{6}, 489 (2010).

\bibitem{Granger}
G.~Granger, J.~P.~Eisenstein, and J.~L.~Reno, Phys. Rev. Lett.
\textbf{102}, 086803 (2009).

\bibitem{Altimiras}
C.~Altimiras, H.~le Sueur, U.~Gennser, A.~Cavanna, D.~Mailly, and
F.~Pierre, Nat. Phys. \textbf{6}, 34 (2010).

\bibitem{G1}
K.~J.~Thomas, J.~T.~Nicholls, M.~Y.~Simmons, M.~Pepper, D.~R.~Mace,
and D.~A.~Ritchie , Phys. Rev. Lett. \textbf{77}, 135 (1996);
K.~J.~Thomas, J.~T.~Nicholls, N.~J.~Appleyard, M.~Y.~Simmons,
M.~Pepper, D.~R.~Mace, W.~R.~Tribe, and D.~A.~Ritchie, Phys. Rev. B
\textbf{58}, 4846 (1998).

\bibitem{G2}
A.~Kristensen, H.~Bruus, A.~E.~Hansen, J.~B.~Jensen, P.~E.~Lindelof,
C.~J.~Marckmann, J.~Nygåard, C.~B.~Sorensen, F.~Beuscher,
A.~Forchel, and M.~Michel, Phys. Rev. B \textbf{62}, 10950 (2000).

\bibitem{G3}
S.~M.~Cronenwett, H.~J.~Lynch, D.~Goldhaber-Gordon,
L.~P.~Kouwenhoven, C.~M.~Marcus, K.~Hirose, N.~S.~Wingreen, and
V.~Umansky, Phys. Rev. Lett. \textbf{88}, 226805 (2002).

\bibitem{G4}
D.~J.~Reilly, G.~R.~Facer, A.~S.~Dzurak, B.~E.~Kane, R.~G.~Clark,
P.~J.~Stiles, R.~G.~Clark, A.~R.~Hamilton, J.~L.~O'Brien,
N.~E.~Lumpkin, L.~N.~Pfeiffer and K.~W.~West, Phys. Rev. B
\textbf{63}, 121311(R) (2001).

\bibitem{G5}
R.~Crook, J.~Prance, K.~J.~Thomas, S.~J.~Chorley, I.~Farrer,
D.~A.~Ritchie, M.~Pepper, C.~G.~Smith, Science \textbf{312}, 1359
(2006).

\bibitem{Chiatti}
O.~Chiatti, J.~T.~Nicholls, Y.~Y.~Proskuryakov, N.~Lumpkin,
I.~Farrer, and D.~A.~Ritchie, Phys. Rev. Lett. \textbf{97}, 056601
(2006).

\bibitem{Wakeham}
N.~Wakeham, A.~F.~Bangura, X.~Xu, J.-F.~Mercure, M.~Greenblatt, and
N.~E.~Husseya, Nat. Commun. \textbf{2}, 396 (2011).

\bibitem{Adilet}
A.~Imambekov and L.~I.~Glazman, Science \textbf{323}, 228 (2009);
A.~Imambekov, T.~L.~Schmidt, and L.~I.~Glazman, Rev. Mod. Phys.
\textbf{84}, 1253 (2012).

\bibitem{KF}
C.~L.~Kane and M.~P.~A.~Fisher, Phyes. Rev. Lett. \textbf{76}, 3192
(1996).

\bibitem{Rosch}
A.~Garg, D.~Rasch, E.~Shimshoni, and A.~Rosch, Phys. Rev. Lett.
\textbf{103}, 096402 (2009).

\bibitem{FHK}
R.~Fazio, F.~W.~J.~Hekking, and D.~E.~Khmelnitskii, Phys. Rev. Lett.
\textbf{80}, 5611 (1998).

\bibitem{Krive}
I.~V.~Krive, Low. Temp. Phys. \textbf{24}, 377 (1998).

\bibitem{GGM}
D.~B.~Gutman, Y.~Gefen, and A.~D.~Mirlin, Phys. Rev. B \textbf{80},
045106 (2009).

\bibitem{MS}
D.~L.~Maslov and M.~Stone, Phys. Rev. B \textbf{52}, R5539 (1995).

\bibitem{GR}
A.~Gramada and M.~E.~Raikh, Phys. Rev. B \textbf{55}, 7673(1997).

\bibitem{Schulz}
H.~J.~Schulz, Phys. Rev. Lett. \textbf{71}, 1864 (1993).

\bibitem{Matveev-WC}
K.~A.~Matveev, Phys. Rev. B \textbf{70}, 245319 (2004).

\bibitem{Bockrath-NatPhys08}
V.~V.~Deshpande and M.~Bockrath, Nat. Phys. \textbf{4}, 314 (2008).

\bibitem{Ritchie-PRL09}
W.~K.~Hew, K.~J.~Thomas, M.~Pepper, I.~Farrer, D.~Anderson,
G.~A.~C.~Jones, and D.~A.~Ritchie, Phys. Rev. Lett. \textbf{102},
056804 (2009).

\bibitem{Yamamoto-PRB12}
M.~Yamamoto, H.~Takagi, M.~Stopa, and S.~Tarucha, Phys. Rev. B
\textbf{85}, 041308(R) (2012).

\bibitem{Matveev-WC-Eq}
K.~A.~Matveev, A.~V.~Andreev, and M.~Pustilnik, Phys. Rev. Lett.
\textbf{105}, 046401 (2010); J.~Lin, K.~A.~Matveev, and
M.~Pustilnik, Phys. Rev. Lett. \textbf{110}, 016401 (2013).

\bibitem{AL-K}
A.~Levchenko, T.~Micklitz, Z.~Ristivojevic, and K.~A.~Matveev, Phys.
Rev. B \textbf{84}, 115447 (2011).

\bibitem{TM}
T.~Micklitz and A.~Levchenko, Phys. Rev. Lett. \textbf{106}, 196402
(2011).

\bibitem{Karzig}
T.~Karzig, G.~Refael, L.~I.~Glazman, and F.~von Oppen, Phys. Rev.
Lett. \textbf{107}, 176403 (2011).

\bibitem{Kapitza-exp}
E.~T.~Swartz and R.~O.~Pohl, Rev. Mod. Phys. \textbf{61}, 605
(1989).

\bibitem{Note}
The dimensionless numerical coefficient in Eq.~\eqref{L-pl} is
$c_\ell=\frac{2100}{\pi^9}\iint^{\infty}_{0}
\frac{x^3y^3(x+y)^3dxdy}{\sinh(x)\sinh(y)\sinh(x+y)}\approx7.61$.

\bibitem{SM}
S.~Apostolov and A.~Levchenko, unpublished.

\bibitem{Iucci}
A.~Iucci and M.~A.~Cazalilla, Phys. Rev. A \textbf{80}, 063619
(2009).

\bibitem{Mitra}
A.~Mitra and T.~Giamarchi, Phys. Rev. B \textbf{85}, 075117 (2012).

\bibitem{Foster}
M.~S.~Foster, T.~C.~Berkelbach, D.~R.~Reichman, and
E.~A.~Yuzbashyan, Phys. Rev. B \textbf{84}, 085146 (2011).

\bibitem{Achim}
A.~Rosch and N.~Andrei, Phys. Rev. Lett. \textbf{85}, 1092 (2000).

\bibitem{Sirker}
J.~Sirker, R.~G.~Pereira, and I.~Affleck, Phys. Rev. B \textbf{83},
035115 (2011).

\bibitem{Roni}
C.~Karrasch, R.~Ilan, and J.~E.~Moore, arXiv:1211.2236.

\bibitem{Polkovnikov}
A.~Polkovnikov, K.~Sengupta, A.~Silva, and M.~Vengalattore, Rev.
Mod. Phys. \textbf{83}, 863 (2011).

\end{thebibliography}
\end{document}